\documentclass[conference]{IEEEtran}
\IEEEoverridecommandlockouts
\usepackage{graphicx} 
\usepackage{float} 
\usepackage{dblfloatfix}
\usepackage[linesnumbered,ruled,vlined]{algorithm2e}
\usepackage{enumitem}
\usepackage{orcidlink}
\usepackage{comment}
\usepackage{hyperref}
\usepackage{amsmath}
\usepackage{eurosym}
\usepackage{subcaption}
\usepackage{csquotes}
\usepackage{doi}
\usepackage[dvipsnames]{xcolor}
\usepackage{xstring}
\usepackage{capt-of}  
\usepackage{cuted}
\usepackage{tikz}
\usepackage{xspace}
\usetikzlibrary{shapes.symbols}
\usetikzlibrary{arrows.meta,bending,positioning}

\newcommand{\ro}[1]{\textbf{RO}$\mathbf{#1}$}
\newcommand{\rd}{\emph{RD}\xspace}
\newcommand{\bd}{\emph{BD}\xspace}

\usepackage{cleveref}
\crefname{section}{Sec.}{Secs.}
\Crefname{section}{Section}{Sections}
\crefname{equation}{Eq.}{Eqs.}
\Crefname{equation}{Equation}{Equations}
\crefname{figure}{Fig.}{Figs.}
\Crefname{figure}{Figure}{Figures}
\crefname{table}{Tab.}{Tabs.}
\Crefname{table}{Table}{Tables}
\crefname{appendix}{App.}{Apps.}
\Crefname{appendix}{Appendix}{Appendix}

\usepackage[backend=biber,
			bibencoding=utf8,
			date=year,
			alldates=year,
            defernumbers=true,
			sorting=none,
			style=numeric,
			citestyle=numeric-comp,
			doi=true,
			giveninits=true,
            urldate=long]{biblatex}

\newcommand{\er}{\mathrm{e}}

\providecommand{\Rcite}[1]{\begingroup
	\def\tempx{0}\StrCount{#1}{,}[\tempx]\ifnum\tempx > 0 
	Refs.~\else
	Ref.~\fi
	\endgroup
	\cite{#1}}

\title{Coin selection by Random Draw according to the 
Boltzmann distribution}

\author{\IEEEauthorblockN{Jan Lennart Bönsel\IEEEauthorrefmark{1}$^{,1}$
\thanks{$^{1}$jan.boensel@bundesbank.de}\orcidlink{0000-0001-6314-945X}, 
Michael Maurer\IEEEauthorrefmark{2}$^{,2}$\thanks{$^{2}$maurer27@mit.edu}\orcidlink{0009-0004-4908-8987}, 
Silvio Petriconi\IEEEauthorrefmark{1}$^{,3}$
\thanks{$^{3}$silvio.petriconi@bundesbank.de}\orcidlink{0000-0003-2638-7525}, 
Andrea Tundis\IEEEauthorrefmark{1}$^{,4}$
\thanks{$^{4}$andrea.tundis@bundesbank.de}\orcidlink{0000-0002-7729-2780}, 
and Marc Winstel\IEEEauthorrefmark{1}$^{,5}$
\thanks{$^{5}$marc.winstel@bundesbank.de}\orcidlink{0000-0003-3470-9926}}
\IEEEauthorblockA{
\IEEEauthorrefmark{1}Deutsche Bundesbank, Wilhelm-Epstein-Straße 14, 60431 Frankfurt am Main, Germany}
\IEEEauthorblockA{\IEEEauthorrefmark{2}MIT Digital Currency Initiative, Cambridge, MA 02139, USA}
\thanks{The views in this paper are those of the authors and not necessarily coincide with those of the Deutsche Bundesbank or the Eurosystem.}
}

\date{\today}

\begin{document}

\maketitle


\begin{abstract}
Coin selection refers to the problem of choosing a set of tokens to fund a transaction in token-based payment systems such as, e.g., cryptocurrencies or central bank digital currencies (CBDCs).
In this paper, we propose the Boltzmann Draw that is a probabilistic algorithm inspired by the principles of statistical physics.
The algorithm relies on drawing tokens according to the Boltzmann distribution, serving as an extension and improvement of the Random Draw method.
Numerical results demonstrate the effectiveness of our method in bounding the number of selected input tokens as well as reducing dust generation and limiting the token pool size in the wallet.
Moreover, the probabilistic algorithm can be implemented efficiently, improves performance and respects privacy requirements -- properties of significant relevance for current token-based technologies.
We compare the Boltzmann draw to both the standard Random Draw and the Greedy algorithm. We argue that the former is superior to the latter in the sense of the above objectives.
Our findings are relevant for token-based technologies, and are also of interest for CBDCs, which as a legal tender possibly needs to handle large transaction volumes at a high frequency.
\end{abstract}



\section{Introduction}\label{sec:intro}


Token-based data models are a key design element in electronic cash proposals \cite{Chaum1983, Brands1994} and cryptocurrencies like Bitcoin
\cite{finney2004rpow, nakamoto2008bitcoin}, and, recently, they are being discussed in the context of central bank digital currencies (CBDCs) \cite{Lee2020,BIS2020}.
We refer to a token as a ``store-of-value" \cite{Lee2020}, i.e., a data object with some immutable value that is transferable by its owner.

Compared to traditional account-based systems, tokens allow for more flexibility in the settlement process \cite{Lee2020}.
Whereas account-based systems require identity verification of the user, a token-based system typically implements algorithms to verify token validity and, potentially, token ownership.
Token verification, however, can be performed on unique tokens independently and it is self-contained, facilitating the parallel settlement of transactions since payments involving different tokens do not conflict.

A prominent example of a token-based model is the Unspent Transaction Output (UTXO) model, which is used in Bitcoin.
In this model, the current system state is defined by the set of all existing UTXOs which have been created but not spent. 
Users spend funds in this system by consuming one or more of these tokens as inputs, and generating new tokens as outputs of the transaction, so long as the transaction is valid.
While Bitcoin is the most prominent example of a UTXO-based model, we stress that tokens do not require a decentralized system. 
For example, \cite{lovejoy2023hamilton} defines a centralized transaction processor based 
on the UTXO model, and additionally demonstrated a throughput of $1.7$ million transactions per second in a 
geographically-distributed payment system under control of a central authority.

Token-based systems, as compared to account-based systems, impose a unique challenge on wallets to select an appropriate subset of tokens (or ``coins'') for a transaction. 
Often this is called the \textit{``coin selection problem''}.
In some scenarios, it is beneficial for a user or system to use a particular coin selection algorithm, such that it may meet certain latency, throughput, privacy, or other goals.
Some users or maintainers of a token-based system may have steep latency and throughput requirements, or strict overall system state size requirements, including for example, commercial banks managing large CBDC wallets \cite{Auer2020}, or popular centralized exchanges aiming to enable real-time settlement on fast moving UTXO-based ledgers. Some examples of such ledgers include Cardano \cite{CardanoDocs}, which has a 20 second blocktime, or Sui, that achieves sub-second settlement times with "Objects" instead of UTXOs \cite{SuiPerformanceUpdate}.

In this context, the coin selection problem involves addressing the following challenges \cite{Ramezan2023,Diroff2019,ChallengesTokenSelection}: (i) \textit{bounding the token pool size in the wallet} -- in order to keep the number of tokens in the wallet, but also the overall
number of tokens in the system, within manageable bounds, e.g., to limit storage requirements.
On the same footing, the selection should (ii) \textit{minimize the number of input 
tokens} to cover a transaction.
Moreover, the algorithm should (iii) \textit{limit the creation of dust}, where dust refers to tokens with low-value which might not be worth spending or simply consume an unjustified amount of resources compared to the value carried.
More general, the wallet should (iv) \textit{preserve a uniform distribution of 
token values} -- in order to reduce disproportionate concentration in certain value ranges to facilitate subsequent transactions.
Finally, the coin selection should be (v) \textit{computationally efficient and 
applicable in concurrent transaction processing}.
This applies in particular to high-throughput scenarios, which for example occur in 
current central bank projects and discussion papers \cite{boe2020discussion,hub2020project, Auer2020}.

From the above mentioned challenges, we formulate the following research objectives (ROs) for the development and evaluation of a coin selection strategy.
The coin selection algorithm should

\begin{itemize}[leftmargin=+1.0cm]
\item[\ro{1}] bound the token pool size,
\item[\ro{2}] limit the number of input tokens for transactions,
\item[\ro{3}] restrict the generation of dust,
\item[\ro{4}] maintain a balanced distribution of token values, and
\item[\ro{5}] handle a high-volume of concurrent transactions. 
\end{itemize}

These objectives, in principle, do not stand independent of each other.
In particular, \ro{3} and \ro{4} aim at keeping the wallet in a healthy state such that they contribute to durably achieving \ro{1}, \ro{2}, and \ro{5} in 
the long run.
Note that many proposals for coin selection algorithms, such as in \cite{Wei2023,Diroff2019,Schneider2024}, motivate \ro{3} by the goal to minimize a transaction fee, which is typically charged based on transaction size in bytes (and other parameters).
In this work, the prevention of dust is further motivated by maintaining a wallet state that facilitates \ro{1} and \ro{2}, while not being wasteful with resources.

To address these objectives, we propose a probabilistic coin selection algorithm, inspired by statistical physics, called the \textit{Boltzmann Draw} (\bd) algorithm.
The \bd is an extension of the \textit{Random Draw} (\rd) that uses the Boltzmann
distribution instead of a uniform probability distribution to sample the tokens.
The single parameter of the Boltzmann distribution is chosen depending both on the total number of tokens in the wallet as well as the sum of their values.
We present findings demonstrating that \bd performs similarly or better than the \rd and the Greedy algorithm with respect to our research objectives.

This article is structured as follows. 
In \cref{sec:relatedwork} we give an overview of coin selection algorithms that
are discussed in the literature. 
After this we introduce the \bd in \cref{sec:algorithm}.
In \cref{sec:analysis}, we present our numerical results comparing the BD to
the \rd and the Greedy algorithm with respect to our objectives.
In \cref{sec:boltzmanndraw_extensions} we point out potential extensions of the  algorithm that could be studied in the future.
Finally, we conclude in \cref{sec:conclusion}.


\section{Related Work analysis}\label{sec:relatedwork}


According to the review in \cite{Ramezan2023}, the coin selection algorithms
in the literature can be classified into three groups:
(i) \textit{Primitive} algorithms mostly utilize elementary properties of 
tokens as age or value, and include, e.g., First In First Out (FIFO), 
Lowest Value First (LVF), and Highest Value First (HVF);
(ii) \textit{Basic} algorithms use some heuristic to find a solution. 
Particular examples are the Greedy algorithm \cite{Wei2023}, Knapsack and Branch \& Bound \cite{Erhardt2016}, as well as probabilistic approaches 
such as \rd or Random Improve \cite{Cardano2025};
finally, (iii) \textit{advanced} algorithms include optimization-based approaches  \cite{Nguyen2018, Abramova2020}, Knapsack with Leverage \cite{Diroff2019}, Greedy and genetic techniques \cite{Holland1992, Wei2023}.

In this section we summarize which objectives the mentioned algorithms address.
For example, LVF reduces dust, whereas HVF minimizes the number of input tokens for
a transaction.
We note that coin selection is a version of the subset-sum problem, which is NP-complete
\cite{Abramova2020,Erhardt2016} and is related to the Knapsack problem \cite{Kellerer2004}.
For these problems, there exist heuristics like the Greedy algorithm \cite{Ramezan2023},
and the Knapsack solver used in Bitcoin \cite{Erhardt2016}.
Bitcoin core wallet also uses Branch and Bound (BnB) \cite{Erhardt2016} and \rd for coin 
selection \cite{BitcoinGithub}.
The Greedy algorithm aims to achieve a minimal number of input tokens (\ro{2}), whereas
Knapsack and BnB try to find exact matches \cite{Ramezan2023}.
Thereby, they avoid change, which reduces the UTXO pool size (\ro{1}).
\rd, in turn, increases the value diversity of the tokens in a wallet and
reveals little information about the wallet state and spending behaviour \cite{Ramezan2023}.

In \cite{Wei2023}, the Greedy algorithm is extended.
The Greedy algorithm is used to generate candidate selections that are further optimized
by a genetic algorithm.
Thereby, this algorithm addresses both \ro{2} and \ro{3}.
The work in \cite{Schneider2024} in turn extends the Greedy algorithm by an adaptive mechanism to
reduce the UTXO pool size.
If the target size of the UTXO pool is exceeded, the algorithm artificially increases the remaining value.
Thus, more tokens have to be selected.
As a result, the algorithm achieves a trade-off between \ro{1} and \ro{2}.

A possible extension of the Knapsack solver is proposed in ~\cite{Diroff2019} with the aim of reducing transaction costs in Bitcoin.
They formulate the Knapsack solver as a binary linear program and add additional
constraints to minimize transaction cost.
The optimization problem is then solved by mixed integer linear programming and 
is referred to as Knapsack with leverage.
If no solution is found, a variant of HVF is used as a fallback.

Ref.~\cite{Ramezan2023b} in turn proposes to optimize coin selection by a mixed integer
non-linear program.
Their multi-asset coin selection (MACS) optimizes for transaction fee, UTXO age, and privacy.
Moreover, they minimize the UTXO pool size (\ro{1}) by limiting the creation of dust (\ro{3}).

Ref.~\cite{Stoyanova2025} combines the BnB algorithm with the DEPS evolutionary solver, which
is a solver for linear optimization problems. 
Compared to the BnB algorithm, the method reduces transaction costs and limits the number
of input tokens per transaction (\ro{2}) by setting an upper bound to the block size.

We conclude that there is a lack of coin selection algorithms taking into account all of the research
objectives, that are identified in \cref{sec:intro}, see also \cref{tab:your_label}.
Moreover, many approaches rely on fallback options such that a single transactions might require
multiple attempts until an appropriate set of tokens is selected.
In addition, optimization algorithms like integer programs can become computationally expensive.

\begin{table*}[h!]
\centering
\begin{tabular}{|c|c|c|c|c|c|c|}
\hline
\textbf{Ref.}  & \ro{1} & \ro{2}& \ro{3} & \ro{4} & 
\ro{5} & \textbf{Adopted approach} \\
\hline
\cite{Ramezan2023}  & - & Yes & - &  - & - & Greedy\\
\hline
\cite{Ramezan2023}  & - & Non expl. & Non expl. &  Yes & Yes & Random  Draw (Probabilistic)\\
\hline
\cite{Diroff2019}  & -  & Non expl. & Yes & - & - & Leveraged Knapsack\\
\hline
\cite{Schneider2024}  & Yes  &  - & Non expl. &  - & - & Adaptive Gen. Alg. \\
\hline
\cite{Ramezan2023b}  & Yes  & - & Yes (age)  & Non expl. & - &  Optimization \\
\hline
\cite{Stoyanova2025}  & -  &  Yes & Non expl. & - & - & BnB + DEPS \\
\hline
\cite{Wei2023}  & - & Yes & Yes &   - & - & Greedy + Gen. Alg.  \\
\hline
\end{tabular}
\caption{
Comparison of the closest related works on coin selection algorithms with respect to our formulated research objectives:
``Yes'' means that the respective objective was explicitly addressed, ``Non expl.'' means that quantitative results are partially neglected or not explained, i.e., the objective is not explicitly addressed, and ``-'' means that the objective is not addressed.}
\label{tab:your_label}
\end{table*}

In contrast, we propose the \bd as a generalization of the \rd in the following section and argue that the \bd achieves 
a reasonable trade-off of the research objectives.
In particular, we show that the \bd is able to handle a high-volume of concurrent transactions.
To the best of our knowledge, concurrency in coin selection algorithms has not been addressed in the literature before.
For this reason, we look to identify algorithms which are "naturally parallel".
Many of the ``basic'' algorithms we've encountered are quite effective for example, but are not amenable to parallelism.
In general, this is because when parallelized, concurrent threads contend on the same ordered data structures (e.g. a global priority-queue when using the LVF algorithm).
This also applies to the Greedy algorithm and BnB \cite{Cardano2025}.
Algorithms like Knapsack and Knapsack with leverage are more friendly to parallelism, since multiple parallel iterations may touch independent data, however they require multiple (dependent) iterations, which increases selection time.

Algorithms like \rd are more naturally parallelizable since concurrent threads are unlikely to select the same token if the total number of tokens is large.
As a probabilistic algorithm, \bd is structurally similar to \rd in the sense that parallel threads are unlikely to select the same token.
Nonetheless, as explained in the following, it might occur that the \bd algorithm creates a bias towards the same section of the shared data structure for all of the concurrent threads, which causes it to be less naturally parallel than \rd. 


\section{The Boltzmann Draw algorithm} \label{sec:algorithm}


\subsection{Motivation}

\begin{figure}[t]
    \centering
    \includegraphics[width=1.0\linewidth]{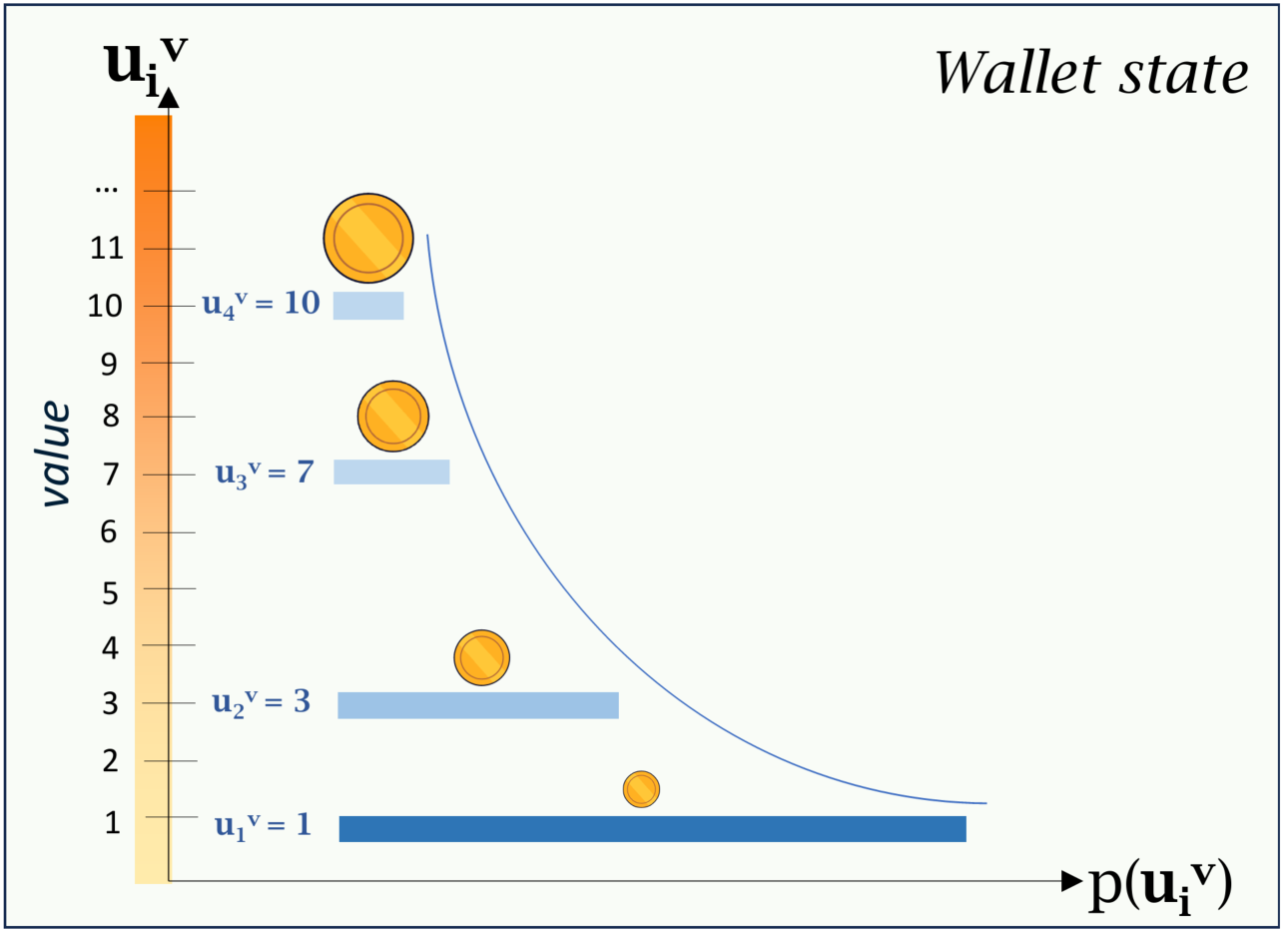}
    \caption{General idea of the \bd selection probability $p(u_i^v)$ of a coin $u_i$ with value $u_i^v$ illustrated on an example wallet with $n=4$ tokens.
    }
    \label{Figure-concepts}
\end{figure}

 
The following proposal for the \bd is inspired by statistical physics and thermodynamics, interpreting the state of a wallet and its tokens in terms of temperature and energy. 
We propose a probabilistic algorithm for coin selection, where each token is picked with a certain probability.
For the selection probability, the value $v$ of a token is interpreted as the energy of a physical state. 
Consequently, the selection probability of a single token is computed according to the Boltzmann distribution occurring in the particle number distribution of classical statistical physics, e.g., in the ideal gas model, see for example \cite{greiner2012thermodynamics} for a generic reference in statistical physics.
Tokens with higher values are, as a result, less likely to be selected. 
The temperature parameter of the Boltzmann distribution is redefined in this context as the average value of tokens in the wallet, see \cref{subsec:boltzmanndraw_algorithm} for details.
A graphical example of the selection probability is represented in \cref{Figure-concepts}. 

A particular inspiration for the BD algorithm stems from the application to high-throughput scenarios, where one wallet must construct many transactions per second and exchanges value with multiple other wallets. 
Using the reinterpretation of value as energy, such a scenario reminds of a thermodynamic system in contact with a large heat bath allowing the system to constantly exchange energy with the heat bath. This only serves as an inspiration for the developed BD algorithm. In particular, this analogy is not a proof or justification of the algorithms effectiveness.

The selection probability of a token mirrors the distribution of particles in real gases or other thermodynamic systems, where states with lower-energy have a higher occupation number (i.e.~probability to be occupied) than higher-energy ones at any finite temperature.
Accordingly, our algorithm is probabilistically designed to preferentially select tokens with lower values -- thereby  directly addressing the issue of dust generation, see \ro{3}.

In the following, we first introduce some basic notation in \cref{subsec:notation} and the BD algorithm itself in \cref{subsec:boltzmanndraw_algorithm}. 


\subsection{Basic notation and Random Draw}\label{subsec:notation}


Let us consider a wallet that contains a pool $U=\{u_{1},\ldots, u_{n}\}$ of $n$ tokens $u_{i}$, $i=1,\ldots,n$.
We denote the value of the \textit{i}th token as $u_{i}^{v}$.
Furthermore, $V$ is the target value of a transaction, i.e., the amount to be paid.
A wallet can make a payment if there exists a subset of tokens $S \subseteq U$  with combined value greater than or equal to $V$, i.e.,
\begin{equation}
    V \text{ can be satisfied} \ \Leftrightarrow \ \exists S \subseteq U : \sum_{u \in S} u^{v} \geq V.
    \label{Basic-constrains}
\end{equation}

Our algorithm can be seen as a generalization of the \rd \cite{Ramezan2023}.
The idea of the \rd algorithm is to randomly extend $S$ by adding tokens from the wallet until the transaction amount is met.
In particular, the token are sampled according to a uniform probability 
distribution.

Let us call $S_{\text{RD}}$ the set of tokens that are selected by the \rd.
We note that the \rd starts with an empty set, i.e., $S_{\text{RD}}=\{\}$. 
$S_{\text{RD}}$ is extended by a random element from $U\setminus S_{\text{RD}}$ as long as the value of the transaction $V$ is not satisfied, i.e., as long as
\begin{equation}
    S_{\text{RD}}^v=\sum_{u\in S_{\text{RD}}}u^{v} < V.
    \label{RandomDrawEquation}
\end{equation}
The tokens in $U\setminus S_{\text{RD}}$ are sampled according to the uniform
probability distribution, i.e.,
\begin{equation}
    p_{\text{RD}}(u) = \frac{1}{|U\setminus S_{\text{RD}}|}
    \qquad\text{for}\qquad u\in U\setminus S_{\text{RD}}. \label{Eq_prob_randomdraw}
\end{equation}
This means that every token has an equal chance of being drawn.

Once $S_{\text{RD} }^v \geq V$, the coin selection algorithm terminates.
The transaction is sent, and the set of tokens $S_{\text{RD}}$ are deleted from
the wallet. 
If the combined value of the set of tokens $S_{\text{RD}}$ exceeds the 
payment amount, a change token is created.


\subsection{The Boltzmann Draw coin selection algorithm}\label{subsec:boltzmanndraw_algorithm}


The proposed \bd algorithm can be understood as a generalization of the \rd.
Instead of using a uniform probability distribution, the tokens are sampled according to the Boltzmann distribution.
Thereby, the algorithm does not only take the number of available tokens into 
account for the selection process, but also their value.
As in the previous case, let us suppose we want to cover a transaction of value $V$ and
denote the set of selected tokens as $S_{\text{BD}}$.
Now, we consecutively add tokens to $S_{\text{BD}}$ until the payment amount $V$ is met.
The tokens are selected at random according to the Boltzmann distribution, i.e.,
\begin{equation}
\begin{split}
    p_{\textrm{BD}}(u)&=\frac{\er^{-\beta(u^{v}-V_{r})}}{\sum_{w\in U\setminus S_{\text{BD}}}
        \er^{-\beta(w^{v}-V_{r})}}\\
        &=\frac{\er^{-\beta u^{v}}}{\sum_{w\in U\setminus S_{\text{BD}}}
        \er^{-\beta w^{v}}}, \qquad\text{for } u\in U\setminus S_{\text{BD}}.
\end{split}
\label{Eq_prob_bolzmann}
\end{equation}
Here, $V_{r}=V-S_{\text{BD}}^{v}$ is the remaining value to be covered, i.e., the difference between target value $V$ and the sum of the token values in $S_{\text{BD}}$. 
The probability in \cref{Eq_prob_bolzmann} 
depends on the value $u^v$ of a token.
Interestingly, the probability distribution is independent of the  remaining target value $V_{r}$.
Note also that the denominator in Eq.~\eqref{Eq_prob_bolzmann} only serves for the normalization of the probabilities to $1$.
The essential behavior of the probability is, thus, given by the exponential function 
\begin{equation}
    w_{\textrm{BD}} (u)= \exp\left( - \beta u^v\right), \label{Eq_Boltzmann_notnormalized}
\end{equation}
which is also referred to as the probability weight of the respective 
token $u$.

The Boltzmann distribution has a free parameter $\beta$. 
As can be deduced from \cref{Eq_prob_bolzmann}, larger values of $\beta$ guide the algorithm to increase the probability of selecting low-value tokens.
This is similar to physical systems where low energy states have higher average occupation numbers when the temperature is lower, i.e., $\beta$ is larger.
As an example how this affects coin selection, let us discuss the probability to select tokens with values $u_1^v = 1$ and $u_2^v = 10$ based on \cref{Eq_Boltzmann_notnormalized}.
For $\beta = 0.1$, one obtains $p_{\textrm{BD}}(u_1)  \approx 0.7109$ and $p_{\textrm{BD}}(u_2) \approx 0.2891 $.
In contrast, for $\beta = 1$ one obtains $p_{\textrm{BD}}(u_1) \approx 0.9999$ and $p_{\textrm{BD}}(u_2)  \approx 10^{-4}$.
Thus, larger positive values of $\beta$ guide the algorithm to select low-value 
tokens with a higher probability, especially compared to the \textit{\rd} algorithm.
Thereby, it directly addresses \ro{3}, i.e., the choice of \cref{Eq_prob_bolzmann} for the selection probability helps to reduce dust. 
In the limit of $\beta=0$, the algorithm behaves identically to the \rd in \cite{Ramezan2023}, since the selection weights are equal for all tokens.

Specifically, one observes that for\footnote{Negative $\beta$ corresponds to temperatures $T< 0$, which can be realized in physical systems in the laboratory. For the sake of the sanity of possible readers we avoid to further elaborate on the physics of ultracold atoms and other laboratory systems achieving negative temperatures.}
\begin{equation}
 \beta \begin{cases}
> 0 & \Rightarrow \textit{low-value tokens } \text{are preferred}, \\
< 0 & \Rightarrow \textit{high-value tokens } \text{are preferred}, \\
= 0 & \Rightarrow \text{Random Draw}.
\end{cases}
\label{Beta_parameter}
\end{equation}

While negative $\beta$ is not a sensible choice with respect to \ro{3}, one still is left with the question of choosing $\beta$.
As demonstrated in the above example computation of probabilities, a sensible choice for the value of $\beta$ is crucial for the performance of the algorithm.
If we draw the analogy to physics, $\beta$ corresponds to the inverse temperature of the wallet, i.e., $\beta \propto 1 / T$, and is related to the entropy.

From the entropy of the wallet we derive in \cref{App_beta} that it is reasonable 
to choose 
\begin{equation}
    \beta=\frac{m}{E},
\end{equation}
where $m$ is the available number of tokens in the wallet for selection, i.e., $m=|U \setminus S_{\text{BD}} |$ with $m \leq n$, and $E$ is the total value of all tokens $u \in U \setminus S_{\text{BD}}$.
This choice is reasonable, since it allows to fix $\beta$ according to the wallet state, in particular through the average value of tokens in the system.

Based on this, given a remaining value $V_{r}$, a random token in the wallet is then picked according to the current probability distribution \eqref{Eq_prob_bolzmann} and added to $S_{\text{BD}}$.
Similar to the \rd algorithm, the coin selection terminates as soon as $S_{\text{BD}}^v \geq V$. 
After each selection of a token, we update $\beta = \frac{m}{E}$ according to the current set of available tokens in $U \setminus S_{\text{BD}}$.
The pseudo-code of the proposed algorithm is given in \cref{alg:pseudo-code}.

\begin{algorithm}
\footnotesize
\caption{ Pseudo-code of the proposed algorithm}
\label{alg:pseudo-code}
\KwData{$TargetPayment$ $V$}
Token-Pool $U = \{u_1, u_2, u_3, \ldots, u_n\}$\; 
 \If{$\sum_{i=1}^{n} u_{i}^{v} < V$}{
  \textbf{return} "Transaction not possible"\;
 }
  \ElseIf{$\sum_{i=1}^{n} u_{i}^{v} = T$}{
     \textbf{return} $U$\;
}
 \Else{
SelectedTokenSet $S_{\text{BD}} = []$\;
TokenChange $u_c^v = null$\;
\While{$(\sum_{u\in S_{\text{BD}}} u^v - V) = u_c^v < 0$}{
TokenWeights $W = []$\; 
$\beta$ = $\frac{|U \setminus S_{\text{BD}}|}{\sum_{u\in U \setminus S_{\text{BD}}} u^{v}}$\;
\For{$i=1, \ldots |U \setminus S_{\text{BD}}|$}{ 
    $W[i]=  w_i \gets w(u_i^v)_{BD}  = \exp\left( - \beta u_i^v\right)$\; 
}
SelectedToken $u \gets getRandomElement(U\setminus S_{\text{BD}}, \mathrm{Weights}=W)$\;
$S.add (u)$\;        
$U.remove(u)$\;

}
\textbf{return} $S_{\text{BD}}$\;
}
\end{algorithm}

Specifically, the algorithm initially checks whether funds are available and terminates if there are not enough (line 2-3). Secondly, it checks whether the value of the tokens is exactly equal to the target value. In this case, it uses all available tokens without any additional calculations (line 4-5).
Otherwise, the selection process begins, iterating until the sum of the values of the selected tokens generates a zero or positive change  (line 9), which means the transaction can be executed.
Specifically, at each iteration—i.e., when selecting a new token—the probability weights \eqref{Eq_Boltzmann_notnormalized} of all unselected tokens are computed. First of all, $\beta$ is calculated (line 11) on the basis of the available number of tokens and their values according to \cref{Beta_parameter} which is then used in the Boltzmann distribution (line 13) according to \cref{Eq_Boltzmann_notnormalized}. 
These probability weights are then used to stochastically select a token from the available token in the wallet (line 14).
The selected token is added to the selected token list and removed from the available remaining tokens (line 15-16).
The algorithm concludes by returning the list of selected tokens along with the change if the total value of the selected tokens is greater than the target value (line 17).


\section{Analysis}\label{sec:analysis}


In this section, we assess the performance of the \bd algorithm.
For this purpose, we simulate the coin selection of a single wallet
and compare the results to the \rd to benchmark against another probabilistic algorithm, and the Greedy algorithm, as a basic algorithm.
\Cref{subsec:scenarios} provides a description of the setup and configuration of the parameters used in the experimental scenarios, while \cref{subsec:results} elaborates on the obtained results in relation to the specific objectives.

\subsection{Scenario description and parameters setting}\label{subsec:scenarios}
The proposed \bd algorithm has been assessed on three test scenarios based on the simulation of deposits (i.e., adding new tokens to the wallet) and payments (i.e., selecting, using, and removing tokens from the wallet).
For the sake of comparison, the first and the second scenarios have been taken from \Rcite{Ramezan2023}. 
Note that only the sender of the transaction must perform coin selection, while deposits produce a single, new token of equivalent value in the wallet.

\begin{figure}[b!]
    \centering
    \includegraphics[width=\linewidth]{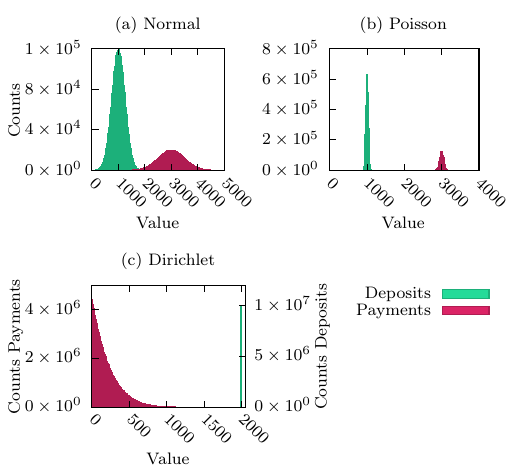}
    \caption{Representation of transaction values under (a) Normal, (b) Poisson, and (c) Dirichlet distributions.} 
    \label{Fig_transaction_distribution}
\end{figure}

\paragraph{First scenario} In the first case, the deposits and payments are assumed to take place according to a \textit{Normal distribution} with mean $1000$ and $3000$, respectively. Furthermore, the deposits have a standard deviation of $250$ and the payments of $500$. In each iteration, $3$ deposits and $1$ payment are simulated. The initial funding of the wallet is one token of value $10^7$.

\paragraph{Second scenario} In the second case, the deposits and payments were chosen according to a \textit{Poisson distribution}, with mean $1000$ and $3000$, respectively. As in the previous case,  $3$ deposits and $1$ payment have been drawn in each iteration. The initial funding of the wallet is one token of value $10^7$.

\paragraph{Third scenario} The third case models the monthly transactions of an individual, who spends all of his monthly deposit. We used the \textit{Dirichlet distribution} assuming $1$ constant deposit of $2000$ and draw $10$ payments, such that their sum matches $2000$.
The wallet is initialized with one token of value $2000$.

The counts of the transaction values is shown in \cref{Fig_transaction_distribution}.
Note that the transaction distribution, i.e., the distribution of payments and deposits, in all scenarios is chosen such that the mean transaction value is zero. 
Nevertheless, in a single simulation one observes a non-vanishing mean transaction value for the first and second scenario, since the mean of the distribution is only achieved in the limit of infinitely many performed transactions.
Thus, the final wallet value after, e.g., $100000$ iterations can fluctuate in these two scenarios. 
This is not the case in the third scenario, where the payments are Dirichlet distributed such that the wallet's total value is unchanged after each iteration.
To ensure that the results are not affected by statistical fluctuations in a single run, many of the following results are averaged over multiple simulation runs (typically $100$) with many iterations (typically $100000$).


\subsection{Results}\label{subsec:results}


The first result concerns the objective \ro{1}, which aims to bound the token pool size within the wallet, i.e., to maintain the total number of tokens at manageable levels.
To assess this, we plot the average token pool size over time.

\begin{figure}[b]
    \centering
    \includegraphics[width=\linewidth]{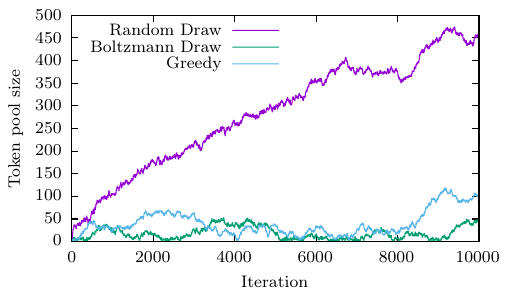}
    \caption{Comparison of one single simulation run with 10000 iterations of the first scenario between \rd, \bd and Greedy.}
    \label{Fig_pool_size_evolution}
\end{figure}

Considering the first scenario, \cref{Fig_pool_size_evolution} illustrates a single simulation run of $10000$ iterations, comparing the results of \rd and Greedy with our proposed \bd approach. An initial observation is the greater efficiency of our algorithm in managing the token pool size compared to \rd, as it consistently maintains fewer tokens in the wallet compared to \rd.
However, note that the token pool size is fluctuating significantly in a single run and token pool sizes achieved by the  \bd and Greedy algorithms are of the same order of magnitude.

\begin{figure*}[pth]
    \includegraphics[width=\linewidth]{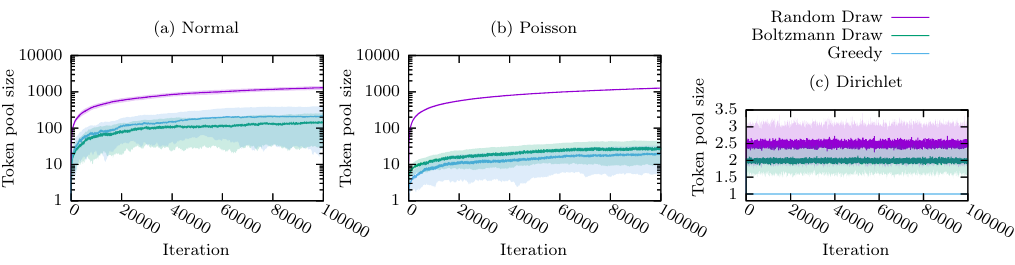}
    \caption{Comparison of the average pool size of 100 simulation runs between \rd, \bd, and Greedy, conducted on the three testing scenarios. The shaded area depicts the standard deviation.}
    \label{Fig_avg_pool_size_evolution}
\end{figure*}

As discussed above, to reduce statistical noise we average over multiple simulation runs.
With respect to token pool size, the result is depicted in \cref{Fig_avg_pool_size_evolution}, where the token pool size is averaged over 100 simulation runs, each consisting of 100.000 iterations.
This demonstrates for three different transaction setups the better performance of the \bd algorithm in bounding the token pool size compared to the \rd algorithm, further validating the finding in the single simulation run from \cref{Fig_pool_size_evolution}.
However, one also obtains that the Greedy algorithm achieves a similar or
slightly smaller token pool size (\ro{1}). 
Nevertheless, in all scenarios the results for Greedy and \bd are within the same order of magnitude, i.e., \bd is competitive with Greedy in \ro{1}.

The most evident difference appears in the second scenario (\cref{Fig_avg_pool_size_evolution}-(b)), where the \bd results in a pool size of almost 30 tokens in the wallet, compared to above 1000 for the \rd, i.e., two orders of magnitude higher. 
Greedy remains in the same order of magnitude as \bd with a pool size of slightly above 20 tokens.
Also in the first scenario (\cref{Fig_avg_pool_size_evolution}-(a)), 
the \bd consistently limits the pool size more effectively than the \rd. 
We note, however, that in the first two scenarios
the number of tokens increases with time.
Finally, in the third scenario (\cref{Fig_avg_pool_size_evolution}-(c)), the three algorithms perform similarly.
The third scenario is a special case because the wallet is initialized with
a token of value $2000$. 
After the deposit in the first iteration there are, thus, two tokens of value $2000$
in the wallet.
With the Greedy algorithm one of those two tokens will be completely spend in the following $10$ payments resulting in a fixed pool size of $1$ token with value $2000$ after the first iteration.
This process repeats itself, such that the Greedy algorithm maintains a fixed pool size of $1$ throughout the whole simulation in this scenario.
The probabilistic nature of \rd and \bd leads to both of the initial two tokens being chosen and, thus, result in slightly higher pool sizes ($2.5$ and $2$, respectively).

\begin{figure*}[bh]
    \centering
    \includegraphics[width=1.8\columnwidth]{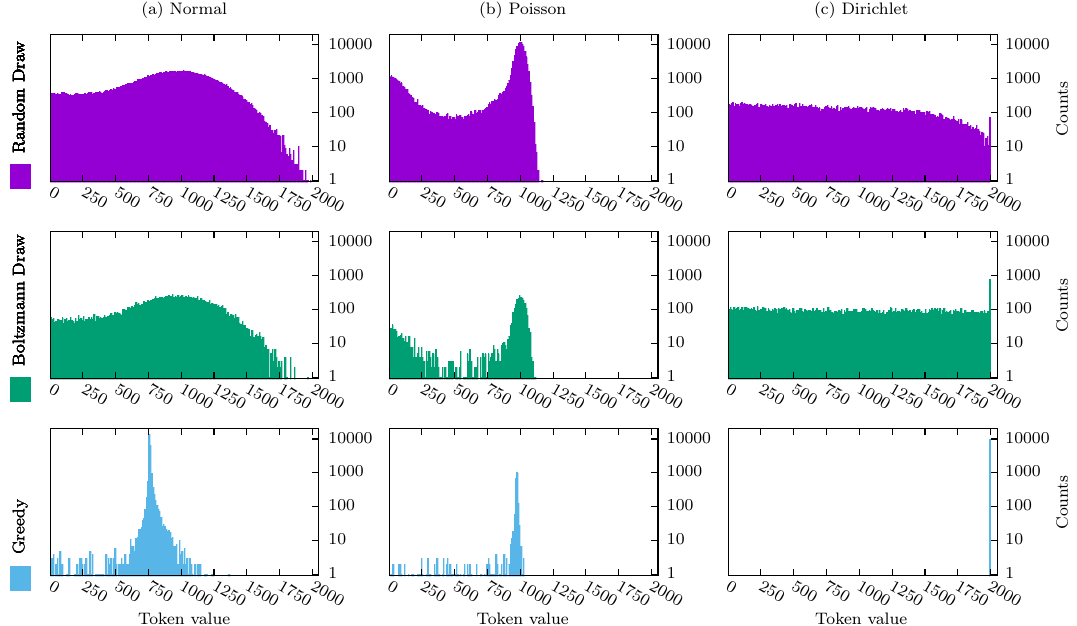}
    \caption{Comparison of the token distributions obtained with the \rd, the proposed \bd, and the Greedy algorithm.
    The data includes the final token counts from multiple simulation runs.
    Thereby, the data for the first and second scenario stems from 100 simulation runs consisting of 100000 iterations each. 
    For the third scenario, the data is obtained from 10000 simulations runs with 1000 iterations each to obtain more statistics because of the small token pool size in a single simulation (see \cref{Fig_avg_pool_size_evolution}).
    In each subfigure, $200$ bins are used.
    Note that the vertical axis is plotted on a logarithmic scale.
    }
    \label{Fig_UTXO_distribution}
\end{figure*}

The second result refers to both objective \ro{3}, that aims at restricting the dust generation (i.e. low-value tokens) as well as objective \ro{4}, i.e., to maintain a balanced diversification of token values in the wallet.
For this purpose, we studied the distribution of tokens in the wallet after the simulation run. 
\Cref{Fig_UTXO_distribution} shows histograms that, for the three algorithms and for each testing scenario, visualize the obtained distribution of tokens at the end of the respective simulation runs.
We note that the token counts are added up from multiple simulations to accumulate more statistics.
Note that the total number of simulation runs differs between the first two scenarios ($100$) and the third scenario ($10000$), since for the third scenario the average token pool size is way smaller, see \cref{Fig_pool_size_evolution}.
For the third scenario, however, the average token pool size does not change already after a few iterations such that only $1000$ iterations are used (compared to $100000$ iterations for the first two scenarios).

Generally, the flatter the histograms in \Cref{Fig_UTXO_distribution}, the more well-distributed the tokens in the wallet, i.e., there is less disproportionate concentration of tokens within any single denomination range (e.g., small, medium, or large), thereby supporting a more versatile set of inputs for subsequent transactions. 
In particular, flat histograms prevent excessive clustering of tokens with similar value, including low-value tokens and consequently reduces the likelihood of creating dust, particularly when encountering an out-of-distribution payment.

\begin{figure*}[t]
    \includegraphics[width=\linewidth]{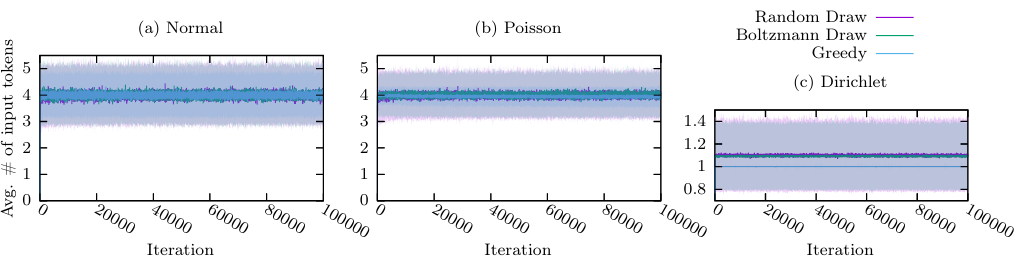}
    \caption{Number of input tokens in each iteration by averaged over $100$ simulation runs of $100000$ iterations each. The shaded area depicts the standard deviation.}
    \label{Fig_avg_number_inputs}
\end{figure*}

As can be seen from this comparison, in each scenario, the curves produced by the two probabilistic algorithms, \rd and \bd, exhibit similar shapes in terms of token distribution, whereas the Greedy algorithm typically exhibits sharper peaks.
In general, the \bd algorithm consistently brings better performance. This can be already observed in \cref{Fig_UTXO_distribution}-(a), corresponding to the first testing scenario, where its curve is lower and smoother in comparison to the classic \rd, with a peak of around $270$ tokens compared to $1771$ (note the logarithmic scale used on the vertical axis of \cref{Fig_UTXO_distribution}), both in bins with token values roughly around $1000$.
The Greedy algorithm exhibits a very sharp peak of $13778$ tokens with values between $745$ and $755$ and, in general, allows for way more clustering of tokens values. 
Specifically, token values above $1000$ are either very sparsely or not at all populated.

In the second scenario (\cref{Fig_UTXO_distribution}-b), the advantage is even more pronounced, with the \bd reaching a peak of about $277$ tokens, whereas the \rd peaks with $12162$. 
Again, the Greedy algorithm exhibits a very sharp distribution, where all bins populated with more than $10$ tokens have values between $925$ and $1005$ and the most populated bin (between values of $965$ and $975$) contains $1012$ tokens.
Similarly, in the third scenario (\cref{Fig_UTXO_distribution}-c), the \bd algorithm generates a flatter curve indicating a more balanced and stable token distribution within the wallet.
This scenario is again a special case, as all final wallet states commonly consist of less than $4$ tokens (see \cref{Fig_avg_pool_size_evolution} for the token pool size).
In a simulation with \bd or \rd, the produced final wallet state either consists of only a single token with a value of $2000$, or two tokens, whose total value is $2000$, as is enforced by the Dirichlet distribution of the transactions.
This is caused by the probabilistic selection of token.
In contrast, simulations with Greedy always maintain a wallet with a single token with value $2000$ after each iteration and, thus, also in the final state of the wallet, see also the discussion of \cref{Fig_avg_pool_size_evolution}.
Averaging over a lot of statistics, the \bd algorithm still produces the most diversified distribution of token values in this scenario compared to both algorithms.

In general, \cref{Fig_UTXO_distribution} suggests that the \bd algorithm generates less tokens with small value (dust) in comparison to \rd for all three test scenarios.
This is in line with its construction outlined in \cref{sec:algorithm}, since the \bd algorithm favors the selection of low value tokens whereas the \rd algorithm  samples tokens independent of their value.
While Greedy typically also produces less dust in absolute numbers, it does not maintain a balanced diversification of token values, thus, not meeting \ro{4}.

In the discussion of \cref{Fig_UTXO_distribution}, it is important to note that the token distributions obtained in \cref{Fig_UTXO_distribution} are certainly also affected by the shape of the drawn deposits, which are independent of the selection algorithm of tokens for payments.
For example, \cref{Fig_UTXO_distribution}-(a) peaks around a token value of roughly $1000$ for \rd and \bd, which is also the mean of the distributions after which deposits are drawn, and \cref{Fig_UTXO_distribution}-(c) shows a significant number of tokens with a value close to $2000$, which is the amount of the deposit in each iteration in the third scenario. 


The third result relates to objective \ro{2} and concerns the number of tokens required for transactions. As in the previous configurations, the experiment compares the proposed solution with the \rd and Greedy algorithms. 
For each of the three test scenarios, the average number of input tokens per iteration was computed by repeating the simulation 100000 times over 100 runs. The corresponding results are presented in \cref{Fig_avg_number_inputs}.
The three algorithms exhibit rather similar performance across the three scenarios.
In the first and second scenario, the average number of tokens per transaction is between $4.0$ and $4.5$ for all algorithms (with minimal difference in fluctuations).
In the third scenario, \rd and \bd converge to an average of approximately $1.1$ tokens per transaction, whereas Greedy averages exactly one token per transaction.
This is in agreement to previous discussions of the Greedy algorithm in context of the third scenario.
Although neither algorithm clearly outperforms the others, all three address the challenge of "minimizing the number of selected tokens" effectively, maintaining a relatively low and nearly constant number of input tokens over time. This consistent behavior, reflected in the close alignment of their results, can be interpreted as a sign of stability in meeting \ro{2} and, consequently, as further evidence supporting the validity of the proposed approach.

\begin{figure}[!b]
    \centering
    \includegraphics[width=1\columnwidth]{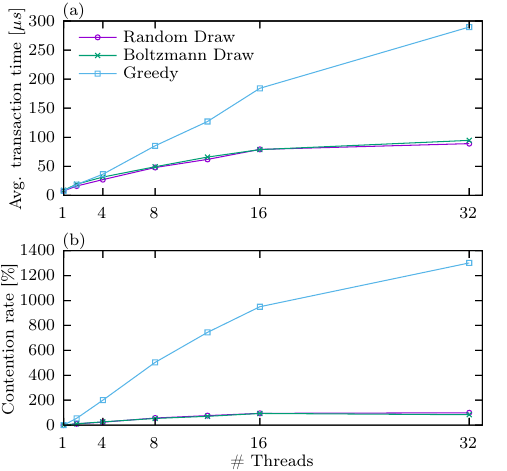}
    \caption{Comparative Latency and Contention Metrics for parallel implementations of \bd (Green), \rd (Purple), Greedy (Light Blue).}
    \label{Fig_parallelism_withoutfifo}
\end{figure}

\textit{Concurrency experiments.}
We run an additional set of experiments aimed at demonstrating the degree of parallelism achievable by each algorithm, and their approximate scaling properties. 
This serves to demonstrate the ability of \bd to address \ro{5} and compare its effectiveness in concurrent applications with other algorithms.
To do this, we implement wallets which support concurrent spending from the same token set, and measure the average time to spend tokens in a single payment as well as the \emph{rate of contention} (i.e. the proportion of "spend" operations which are impeded by a lock at the time of coin selection).
The latter metric allows us to measure the parallelism achievable by the algorithm.
We present results where payments and deposits are generated by a normal distribution as is consistent with the first scenario, see \cref{subsec:scenarios}. The measurements were taken on an Apple MacBook Pro, with a 16-core M3 Max processor, 64 GB of unified memory, running macOS 15.6.1. 
Results are plotted in \cref{Fig_parallelism_withoutfifo}, where we plot the above discussed two metrics as a function of the number of used threads to compare cases with different levels of contention.

\Cref{Fig_parallelism_withoutfifo} demonstrates that the rate of contention and average spend latency increases with the number of concurrent threads go up for each algorithm, as expected, but the effects are most dramatic on the greedy algorithm.
At a high number of threads, the greedy algorithm's performance degrades relative to \bd and \rd because the greedy algorithm has deterministic behavior, and is prefaced on a well ordered data structure. 
In contrast, concurrent threads in \bd and \rd contend infrequently because they are not guaranteed to select the same tokens for a similar spend amount.
Consequently, both \rd and \bd perform similar with respect to both studied metrics.
So, for high-throughput wallets, we find that the two probabilistic algorithms are a better choice compared to greedy, and we contend that this applies to other coin selection algorithms which are deterministic and require a strictly ordered dataset.

\textit{Summary of the achievements:}
Overall, the results clearly show the advantages of the proposed \bd algorithm over the \rd and the Greedy algorithm across multiple dimensions of wallet management with respect to our research objectives for coin selection. In terms of token pool size control (\ro{1}), our approach consistently maintains a significantly smaller and more stable number of tokens in the wallet compared to \rd with reductions reaching an order of magnitude in certain scenarios. 
Between the Greedy algorithm and \bd, token pool sizes are generally of the same order of magnitude. 
Regarding dust prevention (\ro{3}) and token diversification (\ro{4}) \bd achieves a more balanced token value distribution, limiting excessive clustering in specific denominations and, thus, minimizing the risk of low-value token accumulation. Finally, while the three methods perform similarly in minimizing tokens per transaction (\ro{2}), the stability and consistency of the \bd across diverse scenarios further reinforce its robustness. 
Based on the concurrency experiments, we find that \bd has similar latency and contention rate as the \rd, 
whereas the Greedy algorithm has worse concurrency metrics.
These results confirm that our proposal not only addresses the stated objectives effectively, but also provides a more scalable, stable, and dust-resistant solution for long-term wallet management.
Note that, with respect to privacy considerations, probabilistic algorithms, such as \bd and \rd, are preferable over primitive algorithms like Greedy which reveals the range of token values within a wallet to an adversary, that can observe the input tokens of single transaction performed by the respective wallet.


\section{Further ongoing research perspectives}\label{sec:boltzmanndraw_extensions}


In this section, potential ideas about extensions or variants of the above proposed \bd algorithm are highlighted, which in our opinion deserve further attention.
Currently, investigations in these directions have been started. 

\paragraph{Introducing different probability distributions} It is worth noting that the \bd involves the Boltzmann distribution itself only at a single point, when the probability for each token is computed (see \cref{alg:pseudo-code} line 15). 
Thus, a natural variant of the algorithm is a replacement of this probability distribution in line 15 by another function, tuned to fulfill different needs.
Thereby, any function, that produces non-negative weights, can be used.
From a physics perspective, one could, for example, use again a Boltzmann distribution $\er^{-\beta E}$ replacing $E $ with a penalty function encoding certain constraints for token properties, such as their values or ages.
Another simple idea would be using $p(u) = 1 / u^v$, for example. 
This work, to our knowledge, is also one of the only to explore the impacts of different probability distribution based token selection strategies, and it is likely that there are several options in this class of algorithm which may be successful in different scenarios.

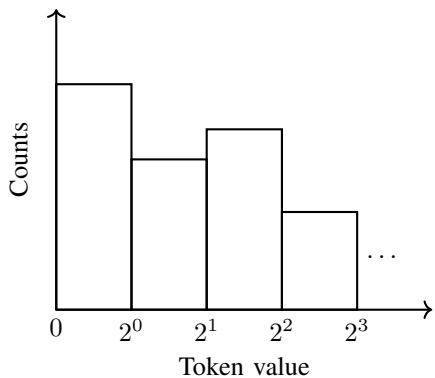
\begin{figure}[h]
    \centering
    \begin{tikzpicture}
        \draw[thick] (0,0) rectangle (1,3);
        \draw[thick] (1,0) rectangle (2,2);
        \draw[thick] (2,0) rectangle (3,2.4);
        \draw[thick] (3,0) rectangle (4,1.3);

        \draw[thick, ->] (0,0) -- (5,0);
        \draw[thick, ->] (0,0) -- (0,4);
        \node[rotate=90] at (-0.5,2) (C) {Counts};

        \node[below] at (0,0) (B0) {$0$};
        \node[below] at (1,0) (B1) {$2^{0}$};
       \node[below] at (2,0) (B2) {$2^{1}$};
        \node[below] at (3,0) (B3) {$2^{2}$};
        \node[below] at (4,0) (B4) {$2^{3}$};
        \node[below] at (2.5,-0.5) (V) {Token value};
        \node[right] at (4,0.7) (Dots) {$\ldots$};
    \end{tikzpicture}
    \caption{Example of a binned token distribution.}
    \label{Fig_binned_distribution}
\end{figure}

\paragraph{Binning token values} Another straightforward idea is to introduce bins of token values.
This reduces the amount of computational resources in the algorithm, which in its original form has to compute probabilities for all tokens in the wallet.
This becomes costly whenever token pool sizes are large.
Thus, we propose to bin tokens by their value, for example covering powers of $2$, e.g. $B_{0}=[0,1)$ and $B_i=[2^{i-1}, 2^{i})$ for $i\geq 1$; or powers of $10$, e.g.,  $B_0=[0, 10^{-2})$ and $B_j = [10^{-2+j - 1}, 10^{-2+j})$ for $j\geq 1$. 
\Cref{Fig_binned_distribution} shows an example of a binned token 
distribution.
In the case of binning, one possibility to compute the probability of selecting a certain bin is
\begin{equation}
    p(B_j) = \begin{cases} \exp\left(- \beta \sup B_j\right) & \mbox{, if } B_j \mbox{ is filled;} \\  0 & \mbox{, otherwise} \end{cases} . 
\end{equation}
Here, $\sup B_j$ denotes the supremum of the respective bin interval $B_j$, i.e., its upper boundary.
A bin $B_j$ is considered as filled, whenever there exits at least one token $u_k \in U$ with $u_k^v \in B_j$. 
Once a bin is selected, tokens within the selected bin can be chosen using the \rd algorithm.
The binning can in a later stage also be dynamically changed to address particular states of the wallet such as, e.g., a low token pool size with a large proportion of high value tokens.

\paragraph{Introducing chemical potential(s) for bins}
In statistical physics, a Boltzmann distribution describes the average particle occupation numbers in a system allowed to exchange energy with a large heat bath, i.e., a system with fixed temperature and particle number, as can be derived from the canonical ensemble formalism.
There is a natural extension to open systems, where also particle numbers can fluctuate: the grand-canonical ensemble, where a so-called chemical potential $\mu$ is introduced to fix an average total particle number. 
With respect to the Boltzmann distribution as selection probability \eqref{Eq_Boltzmann_notnormalized}, this introduction amounts to replacing $\er^{- \beta u^v} \rightarrow \er^{-\beta (u^v - \mu)}$.
Similar to the remaining transaction value $V_r$ in \cref{Eq_prob_bolzmann}, the chemical potential $\mu$ would, however, drop out the probability distribution function after normalization. 
However, one could introduce bins $B_j$ in the sense of the preceding paragraph and introduce a chemical potential $\mu_j$ for each bin, which depends on the number of tokens with values within the respective bin interval. 
Such an algorithm could allow to purposefully influence the ``filling level'' of bins and, thereby, the value distribution.


\section{Conclusion}\label{sec:conclusion}


The paper proposes a novel coin selection strategy, which focuses on optimizing five key objectives: \ro{1} bounding the token pool size within the wallet, \ro{2} limiting the number of tokens involved in the transactions, \ro{3} mitigating the accumulation of low-value tokens (dust), \ro{4} preserving a balanced token value diversification and \ro{5} being effective in concurrent transaction processing.

By creating the analogy of the value of a token to the energy of a particle in a thermodynamic system, an \textit{ad-hoc} coin selection algorithm, called \textit{Boltzmann Draw} (\bd), is defined using the Boltzmann distribution as the probability distribution to select tokens. 
The Boltzmann distribution leads to higher probabilities for low value tokens and lower selection probabilities for high value tokens, as compared to a uniform probability distribution which is used in the standard \textit{Random Draw} (\rd) algorithm.
 
Numerical simulations have been conducted by simulating the automatic selection of tokens in a user wallet employing three scenarios of transactions.
Thereby, deposits and payments are generated according to different distributions (i.e. Normal, Poisson, and Dirichlet).
Moreover, experiments with realistic hardware are performed to test for efficiency in concurrent workloads (cf.~\ro{5}). 
The obtained results are benchmarked against the \rd and the Greedy algorithm.
Our numerical results suggest that the \bd algorithm outperforms the \rd algorithm in \ro{1}, \ro{3} and \ro{4}, whereas similar performance of both algorithms is observed in \ro{2} and \ro{5}.
Comparing to the Greedy algorithm, \bd performs similarly with respect to \ro{1} and \ro{2}, and provides better performance with respect to \ro{3}, \ro{4} and \ro{5}. 
In general, we expect that the probabilistic \bd and \rd algorithms achieve better contention metrics compared to Greedy for highly threaded applications.
The \bd algorithm is (computationally) cost efficient compared to optimization based approaches and also allows for application in high-throughput scenarios, where concurrent processing of transaction can be beneficial. The latter is not easily achievable in optimization-based approaches.
As a probabilistic selection algorithm, the wallet fingerprint is also less well-defined, which is advantageous for privacy in most scenarios. 
This is an advantage of \rd and \bd over the Greedy algorithm, that potentially allows to infer information about the range of values stored in a respective wallet.
Even when observing all input tokens of one or more transactions performed by a wallet applying the \bd selection method, an adversary can only gain information about the average values of tokens in the wallet by reconstructing the probability distribution.\footnote{Note that the average value of tokens and, thereby, the $\beta$ parameter of the \bd changes with each transaction, such that the knowledge of multiple transactions of a single wallet might not necessarily help an adversary in gaining this information.} 

The proposed algorithm can be adapted by changing the interpretation of energy in the Boltzmann distribution and incorporating other token properties, such as age.
Moreover, the application of (possibly dynamic) binning of token values can improve its
efficiency further.
Another possible venue of application, which is to be explored, is the behavior of the algorithm in an explicit high-throughput scenario with concurrent processing of transactions.
These and other further research directions are interesting and currently under pursue of the authors.


\section*{Acknowledgment}


We acknowledge useful discussions with Neha Narula. 
M.W.~acknowledges support by the Giersch Foundation.


\printbibliography

\appendix


\subsection{Partition function of the wallet and $\beta$}\label{App_beta}


It is sensible to choose the parameter $\beta$ of the Boltzmann distribution, used as the major component of the \bd algorithm, presented in \cref{sec:algorithm}, dynamically according to the state of the wallet.
Therefore, we calculate the partition function $Z$ of a wallet with total value $E$. 
\begin{equation}
\begin{split}
    Z&=\sum_{U^{v}\leq E}1
    =\mathcal{N}\int_{0}^{E}du_{1}^{v}\int_{0}^{E-u_{1}^{v}}du_{2}^{v}
        \ldots\int_{0}^{E-\sum_{j=1}^{n-1}u_{j}^{v}}du_{n}^{v}\\
    &=\mathcal{N}\frac{E^{n}}{n!}
\end{split}
\end{equation}
with an irrelevant normalisation constant $\mathcal{N}$.
In the above evaluation, we have made use of the volume of an $n$-simplex.
We can now calculate the entropy $S$ of the wallet:
\begin{equation}
\begin{split}
    S&=k_{B}\ln\left(Z\right)
    =k_{B}\ln\left(\mathcal{N}\frac{E^{n}}{n!}\right)\\
    &=k_{B}\left[\ln\left(\frac{\mathcal{N}}{n!}\right)+\ln(E^{n})\right]
    =k_{B}\left[\ln\left(\frac{\mathcal{N}}{n!}\right)+n\ln(E)\right].
\end{split}
\end{equation}
From a physics context, this yields for the parameter $\beta$:
\begin{equation}
   \beta=\frac{1}{k_{B}T}=\frac{1}{k_{B}}\frac{\partial S}{\partial E}
    =\frac{n}{E}.
\end{equation}
This is good for privacy, since the distribution only contains information about the average value of the tokens.


\end{document}